\documentstyle[graphics,psfig]{aa}

\begin{document}

\title{XMM-Newton observation of the Seyfert 1.8 ESO\,113-G010: 
discovery of a highly redshifted iron line at 5.4\,keV. 
}

\author{D. Porquet\inst{1} \and J.N. Reeves\inst{2,3} \and P. Uttley\inst{2} \and T.J. Turner\inst{2,4}}
\offprints{D. Porquet}
\mail{dporquet@mpe.mpg.de}

\institute{
Max-Planck-Institut f\"{u}r extraterrestrische Physik, Postfach 1312,
 85741 Garching, Germany
\and Laboratory for High Energy Astrophysics, 
NASA Goddard Space Flight Center, Greenbelt, MD 20771, USA
\and Universities Space Research Association, 7501 Forbes Boulevard, Suite 
206, Seabrook, MD 20706, USA
\and Joint Center for Astrophysics, Physics Department, 
University of Maryland Baltimore County, 1000 Hilltop Circle, 
Baltimore, MD 21250, USA  
}

\date{Received June 2, 2004 / Accepted July 17, 2004}

\abstract{We present a spectral analysis of the Seyfert 1.8  
\object{ESO 113-G010} observed with {\sl XMM-Newton} for 4\,ks. 
 The spectrum shows a soft excess below 0.7\,keV and more interestingly 
a narrow emission Gaussian line at 5.4 keV (in its rest-frame), most probably 
originating from a redshifted iron K$\alpha$ line. 
No significant line at or above 6.4\,keV is found contrary to other 
objects showing redshifted lines, ruling out a strong blue-wing to the 
line profile.  The line is detected at 99\% confidence, from performing 
Monte Carlo simulations which fully account for the range of energies 
where a narrow iron line is likely to occur.  
The energy of the line could indicate emission from 
relativistic (0.17--0.23\,c) ejected matter moving away from the observer,  
as proposed for Mrk 766 by Turner et al. (\cite{Tu04}). 
Alternatively, the emission from a narrow annulus 
 at the surface of the accretion disk 
is unlikely due to the very small inclination angle 
 (i.e. less than 10$^{\circ}$) required to explain the narrow, 
redshifted line in this intermediate Seyfert galaxy. 
However emission from a small, localized 
hot-spot on the disk, occurring within a fraction of a complete disk orbit,  
could also explain the redshifted line. 
This scenario would be directly testable in a longer 
observation, as one would see significant variations in the energy and 
intensity of the line within an orbital timescale. 

\keywords{galaxies: active - galaxies: quasars - X-rays - 
Line: formation - X-rays: individual: ESO 113-G010 }  
}
\titlerunning{XMM-Newton observation of ESO 113-G010 }
\authorrunning{Porquet et al.}
\maketitle
\section{Introduction}

In Active Galactic Nuclei (AGN), from Seyfert galaxies to quasars,
 the analysis of several X-ray features can help us to understand
 the central region of these powerful objects. 
 Especially in the hard X-ray band,
the Fe\,K${\alpha}$ line complex observed in the 6--7\,keV range is
an important spectral diagnostic tool to probe dense matter
 from the inner disk out to the Broad Line Region and the molecular torus
 (see review in Reynolds \& Nowak \cite{RN03}).
Recently, narrow spectral features in the 5--6\,keV energy range 
were discovered with {\sl XMM-Newton} and {\sl Chandra} in a few AGN:  
\object{NGC 3516} (Turner et al. \cite{Tu02}), 
\object{ESO 198-G24} (Guainazzi \cite{Gu03}, Bianchi et al. \cite{B04}), 
\object{NGC 7314} (Yaqoob et al. \cite{Y03}), 
\object{Mrk 766} (Turner et al. \cite{Tu04}), 
and  \object{IC\,4329A} (McKernan \& Yaqoob \cite{MY04}).  
Localized spots or narrow annuli which occur on the surface of an accretion disk 
following its illumination by flares has been proposed to explain these features
 (e.g., Nayakshin \& Kazanas \cite{NK01}, Turner et al. \cite{Tu02}, Dov\v{c}iak et al. \cite{D04}).
An alternative scenario has been proposed by Turner et al. (\cite{Tu04})
 for Mrk\,766 for which the first 
evidence for a significant line energy shift has been observed over a few tens
 of ks. They proposed that this shift could be 
interpreted as deceleration of an ejected blob of gas traveling close to 
the escape velocity. \\

Here, we present the first observation above 2\,keV  by {\sl XMM-Newton}
of the Seyfert type 1.8 galaxy ESO\,113-G010 
(z=0.0257, FWHM ${\rm H\alpha}$ = 2\,000\,km\,s$^{-1}$,
 Pietsch et al. \cite{Pi98}). 
The present {\sl XMM-Newton} data shows that 
this object shows a soft excess and more interestingly 
an emission feature near 5.4 keV (in the Seyfert rest-frame). 
H$_{\rm 0}$=75\,km\,s$^{-1}$\,Mpc$^{-1}$  
and q$_{\rm 0}$=0.5 are assumed throughout.

\section{XMM-Newton observation}\label{sec:xmm}

ESO\,113-G010 was observed by XMM-Newton on May 3, 2001
 (OBSID:0103861601).  
The EPIC MOS cameras (Turner et al. \cite{Turner2001}) 
were operating in partial window mode (exposure time about 6.9\,ks), 
and the EPIC PN camera (Str\"uder et al. \cite{Strueder2001})
 was operating in full window mode (exposure time about 4.1\,ks).
The data were re-processed and cleaned using the 
{\sl XMM-Newton} {\sc SAS version 5.4.1} (Science Analysis Software) package. 
 Data are selected using event patterns 0--4 and 0--12 for pn and MOS,
 respectively, whilst only good X-ray events (with 'FLAG=0') are included. 
 The source spectra are extracted from a circular source region 
of about 32.6$^{\prime\prime}$ radius,
 and the background spectra are extracted from a box ($6^{\prime}\times 3^{\prime}$)
 close to the source.
The pn net exposure time is 4\,ks. 
The $S/N$ of the MOS data is too low above 5\,keV, 
therefore we use only the PN data which has the greatest sensitivity in the 
iron K-shell band. 
Unfortunately, for the present exposure time the
 $S/N$ of the RGS data are by far too low for any spectral analysis. 

\section{Spectral analysis}\label{sec:xmmspectra}

In the following analysis, abundances are set to 
those of Anders \& Grevesse (\cite{Anders89}), whilst 
we use the updated cross-sections for X-ray absorption by 
the interstellar medium  from Wilms et al. (\cite{Wilms2000}).  
 The Galactic column density value is 2.74 $\times$ 10$^{20}$\,cm$^{-2}$, 
 obtained from the {\sc coldens} program using the compilations of Dickey \& Lockman (\cite{DL90}). 
We bin the background-subtracted spectrum in order to have a minimum
 of 20 counts per bin. The errors quoted correspond to 90$\%$ 
confidence ranges for one interesting parameter 
($\Delta \chi^{2}$=2.71).\\

\begin{figure}[t!]
\psfig{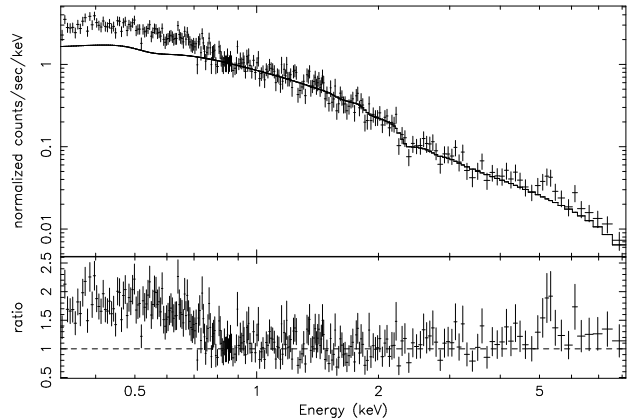}
 \caption{PN spectrum of ESO 113-G010 (z=0.0257, observer frame). 
A power law has been fitted to the 1--4 keV data and extrapolated
 to lower and\ higher energies. A soft X-ray excess is clearly seen extending to about
 0.7\,keV, as well as a positive deviation near 5.4 keV.}
 \label{fig:spec}
 \end{figure}

First we fit the data in the 1--4\,keV energy range where the spectrum 
 is relatively unaffected by the presence of a soft excess below about
 0.7\,keV, by a possible warm absorber, 
and by a Fe K emission line.
 Over this energy range, the data are well fitted by a single power law model
 with $\Gamma$=2.07$\pm$0.08 ($\chi^{2}$=113.4/116). 
Figure~\ref{fig:spec} displays the spectrum extrapolated 
over the 0.3--8 keV broad band energy. 
A strong positive residual is clearly seen below about 
0.7 keV due to the presence of a soft X-ray excess.
 In addition, a clear positive deviation near
 5.4 keV (in the Seyfert rest-frame) is observed.

\subsection{The soft X-ray spectrum}

The spectrum below 2\,keV is only moderately 
fitted  ($\chi^{2}$/dof=268.5/197) 
by an absorbed (by Galactic absorption and intrinsic absorption) 
power law  with $\Gamma$=2.67$^{+0.08}_{-0.05}$
 (due mainly to the presence of a soft excess below 0.7\,keV). 
We find an upper limit for intrinsic absorption at the redshift 
of the Seyfert of 0.8 $\times$ 10$^{20}$\,cm$^{-2}$.
Since this value is much weaker than the Galactic absorption, 
we fix the absorption column density to that
of the Galaxy (i.e., 2.74 $\times$ 10$^{20}$\,cm$^{-2}$).  
We inferred a 0.1--2.4\,keV luminosity of about
 1.4 $\times$ 10$^{43}$\,erg\,s$^{-1}$, which is compatible
with the previous observations with {\sl ROSAT} (Pietsch et al. \cite{Pi98}).  
The standard explanation for the soft X-ray excess is that it results
  from thermal emission originating directly from the hot inner accretion disk  
(Malkan \& Sargent \cite{MS82}), 
and hence it is the high energy tail of the so-called Big Blue Bump.  
 We fit the data with the combination of an accretion disk black body spectrum 
  ({\sc diskpn} in XSPEC, Gierli{\' n}ski et al. \cite{Gi99})   
 and a power law continuum, for the soft and hard band, respectively.  
We find a rather good fit with kT=121$^{+16}_{-10}$\,eV 
and $\Gamma$=1.94$\pm$0.07 ($\chi^{2}$/d.o.f=267.9/250).

\subsection{The spectral feature at 5.4 keV}

 \begin{figure}[t!]
 \psfig{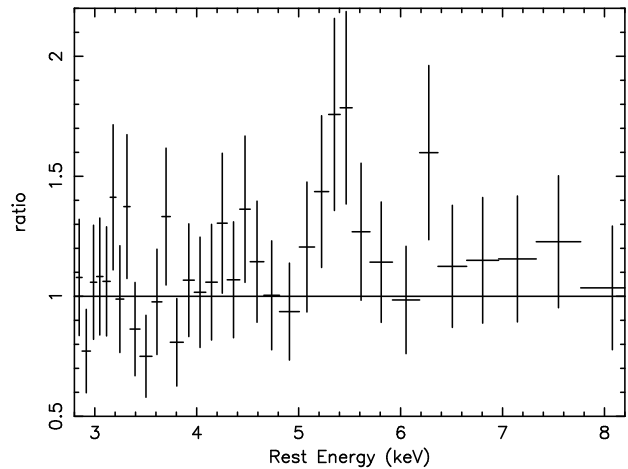}
 \caption{A close-up of the data/model ratio in the 3--8\,keV energy range.
The model is a Galactic absorbed power law. A clear positive residual 
is seen near 5.4\,keV.} 
 \label{fig:ratio}
 \end{figure}
\begin{figure}[t!]
\begin{tabular}{c}
 \psfig{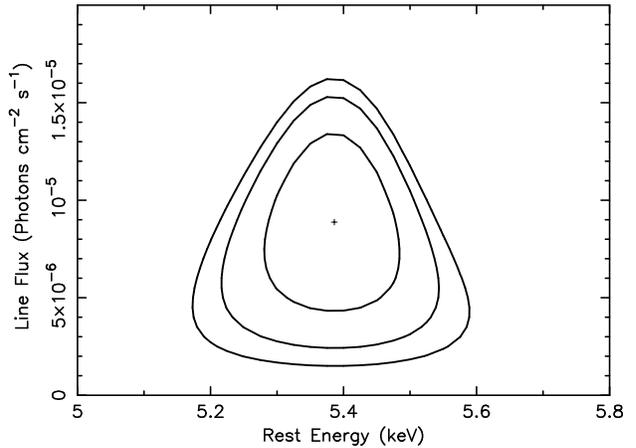}\\
\end{tabular}
 \caption{Contour plot showing the 68$\%$, 90$\%$ and 95$\%$
 (from the inner to outer curves) confidence level for 
values of line rest-energy (keV) and line flux for the line detected near 5.4\,keV.}
 \label{fig:contour}
 \end{figure}

Fitting the data above 1 keV with a 
(Galactic) absorbed power law gives a good fit with $\Gamma$=1.98$\pm$0.06 
($\chi^{2}$/d.o.f=125.1/128). The inferred unabsorbed 2--10\,keV flux 
 is 2.4 $\times$ 10$^{-12}$\,erg\,cm$^{-2}$\,s$^{-1}$,
 corresponding to a luminosity of 3.1 $\times$ 10$^{42}$\,erg\,s$^{-1}$. 
There is a clear positive residual near 5.4 keV as also shown
 in Figure~\ref{fig:ratio}. The feature is well modeled by a Gaussian line 
(with $\sigma$=0.1\,keV) with $\Delta \chi^2$=8.7 for 2 additional parameters. 
 The line energy is 5.38$\pm$0.11\,keV with EW=265$^{+147}_{-145}$\,eV.   
 Figure~\ref{fig:contour} shows the confidence contour plot for the rest-energy
 and intensity of the line corresponding to the 
 68$\%$, 90$\%$ and 95$\%$ confidence levels. 
 One should notice that a Cr K${\alpha}$ (at 5.4\,keV) is present in the internal EPIC-pn background 
 (see Figure~2 in Freyberg et al. \cite{F02}) 
 but its strength is negligible and therefore can be excluded 
as a possibility to explain the 5.4\,keV line.  
Formally, we do not find any statistical requirement for a narrow line 
at 6.4\,keV ($\Delta \chi^2$=0.8 for 1 additional parameter).
Fixing the line energy at 6.4~keV and the width at 10\,eV 
(i.e an intrinsically narrow line), 
we find an upper limit for the EW of 230\,eV.
Splitting the observation into two equal parts (of about 2\,ks each),
 we do not find any significant line parameter variation 
 during this very short timescale. \\

\subsection{Estimating the detection probability of the line}

\indent In estimating the significance of the 5.4\,keV line, 
straight-forward application of the standard two parameter 
F-test does not take into 
account the range of energies where lines can occur in the spectrum, 
nor does it easily account for the number of bins in which one is 
searching for lines. The probability $P_{0}$ of finding a feature at one 
particular given energy can be estimated from the F-test,
 upon the addition of one extra degree of freedom (the line normalisation),
 which in ESO\,113-G010 is 99.8$\%$.
 However in searching for redshifted (or blueshifted) iron 
lines, we are fitting over many bins or resolution elements, where 
narrow features may occur by chance (i.e. statistical noise). 
Thus the probability of detecting a line at {\it any} energy in the iron 
K-shell band can be estimated by $P_{N}=P_{0}^N$, 
where N is the number of bins searched. In ESO\,113-G010, we define the 
iron K band between 4--7 keV (i.e. where one may realistically expect to 
find iron lines), which consists of 18 spectral bins (at 20 counts 
per bin). Thus the probability of detection of the 5.4 keV feature by this 
estimate is 96.5\%.  Note this is probably an under-estimate of the 
detection probability, as some of the actual 
spectral bins are smaller than the 
instrument resolution and therefore the number of bins (or trials) where 
a line may be located has probably been over-estimated. \\

\begin{figure}[t!]
\begin{tabular}{c}
\psfig{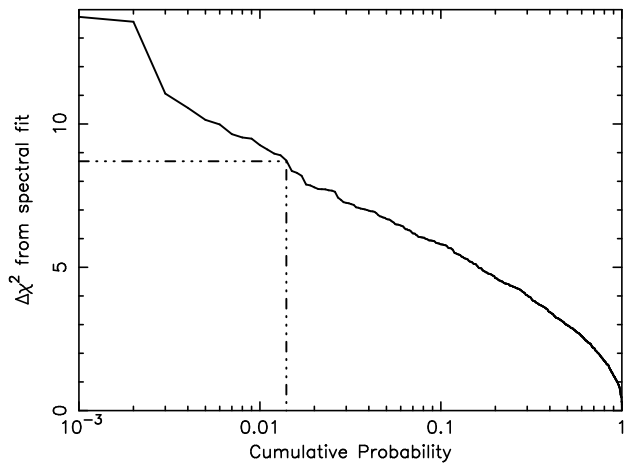}
\end{tabular}
\caption{The measured change in $\chi^{2}$ against cumulative probability 
for the addition of a single Gaussian line in 1000 randomly generated 
simulated spectra. The null hypothesis for the simulated spectra is 
a power-law continuum with Galactic absorption. For 
ESO\,113-G010, the measured $\Delta \chi^{2}=8.7$ for the addition of 
a Gaussian line at 5.4\,keV is only exceeded 
in 1.4\% of the simulated spectra. 
Therefore the 5.4 keV line is detected at $\sim99$\% confidence.} 
\label{fig:montecarlo}
\end{figure}

 We also carried out a more rigorous test of the significance of the line
using Monte Carlo simulations.  For our null hypothesis, we assumed
that the spectrum is simply an absorbed power-law continuum, with the same
parameters as the absorbed power-law model fitted to the real data.
We used the {\sc xspec} {\sc fakeit} command to create 1000 fake 
EPIC-pn spectra corresponding to this model, with photon statistics expected
from the 4~ks exposure, and grouped each spectrum to a maximum
of 20 counts per bin.  Following the procedure used to test the real
data for the presence of a narrow line, we fitted each fake spectrum with
an absorbed power-law (absorption fixed at Galactic, but power-law 
photon index and normalisation left free to vary), to obtain a $\chi^{2}$
value.  We then added a narrow line ($\sigma=0.1$~keV) to the fit,
restricting the line energy to be between 4--7~keV. Furthermore
we stepped the line over the 4--7~keV energy range in increments of 0.1~keV, 
whilst fitting separately each time to ensure the lowest $\chi^{2}$ value 
was found.  We then recorded the minimum $\chi^{2}$ obtained
from these multiple line fits for each fake spectrum, and compared
with the corresponding $\chi^{2}$ of the null hypothesis fits, to obtain
1000 simulated values of the $\Delta \chi^{2}$, which we used
to construct a cumulative frequency distribution of the $\Delta \chi^{2}$
expected for a blind line search in the 4--7~keV range, assuming the null
hypothesis of a simple power-law with no line is correct. 
The resulting distribution of $\chi^2$ against detection probability is shown in 
Figure~\ref{fig:montecarlo}. We therefore find only
1.4\% of fake power-law spectra fitted with a line
show a larger $\Delta \chi^{2}$ than observed in the real data,
implying that the line detection is significant at 
approximately 99\% confidence.

 \begin{figure}[t!]
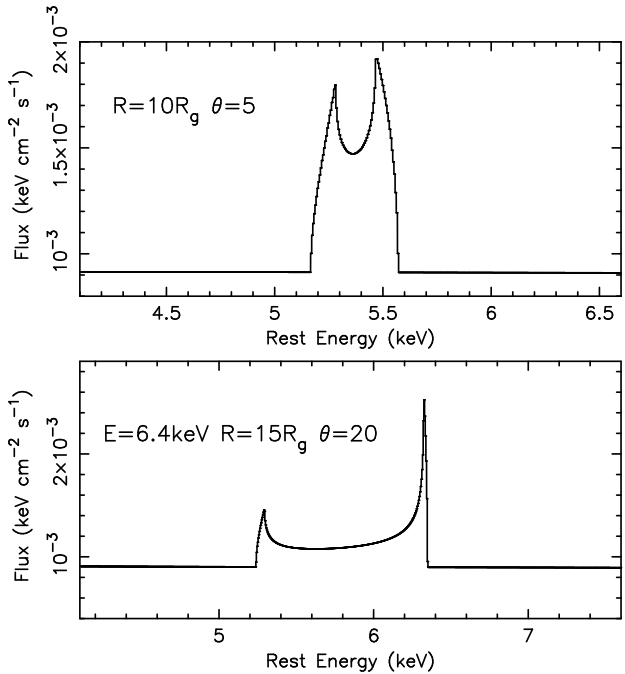

\begin{tabular}{c}
 \psfig{file=1637_f5a.ps,width=8.2cm,angle=-90}\\
 \psfig{file=1637_f5b.ps,width=8.2cm,angle=-90}\\
\end{tabular}
\caption{Line profiles obtained by a relativistic line model 
 for two inclination angles ($\theta$). 
Top: $\theta$=5$^{\circ}$. Bottom: $\theta$=20$^{\circ}$.} 
\label{fig:angle}
 \end{figure}

\subsection{A relativistic disk origin for the line?}

\indent The 5.4~keV line could be a relativistic line for which we only see 
 one horn due to the moderate spectral resolution.
 Therefore, we fit the data with a {\sc diskline} profile
 (Fabian et al. \cite{F89}). We assume a narrow annulus  
of 1 R$_{\rm g}$, in order to obtain a narrow profile.
Initially we fix the inclination angle at
 $\theta=45^{\circ}$, and assume the line is emitted at 6.4~keV 
(i.e near-neutral iron). 
 However this model fails to re-produce the narrow emission line at 
5.4 keV. Upon varying the disk inclination angle a good fit to the line is 
obtained, with $\chi^{2}$/d.o.f=118.4/127, 
requiring a very small inclination angle of $\theta\sim5^{\circ}$, 
with a formal upper-limit at 90\% confidence of $\theta<10^{\circ}$. 
We constrain the inner radius of the annulus at
 R$_{\rm in}$=9.5$^{+1.1}_{-0.9}$\,R${\rm g}$, 
whilst the line EW is 303$\pm$166\,eV. 
Thus for a very low (face-on) disk inclination  
the line is gravitationally redshifted whilst there is very
little transverse velocity shift, hence the red and blue wings of the line are 
not resolved. A narrow redshifted feature is then observed, consistent 
with the line profile in Figure~2. 
At higher disk inclination angles, the transverse velocity shift 
is much more prominent leading to a strong blue wing, which is not 
observed in the data. 
This is illustrated in Figure~\ref{fig:angle}
where the theoretical line profiles at two angle inclinations are compared, 
i.e., $\theta=$5$^{\circ}$ and $\theta$=20$^{\circ}$.

\section{Discussion}

It has been proposed by Skibo (\cite{Sk97}) that energetic protons
 can be responsible of the destruction (spallation) of Fe into Cr and
 other lower $Z$ metals on the surface of the accretion disk, which enhances the
 line emission expected from elements of low abundance.
The strongest line apart the Fe\,K$\alpha$ line expected is the Cr\,K$\alpha$
 line at 5.4\,keV, and the second one being the Mn\,K$\alpha$ at 5.9\,keV
 (see Figure 4 in Skibo \cite{Sk97}).
The energy of the feature we find here is compatible with the energy of the
 Cr\,K$\alpha$ line. 
However the ratio of the line at 5.4\,keV (if Cr\,K${\alpha}$) 
and the upper limit of a narrow Gaussian at 6.4 keV (due to iron) is not
 consistent with the line ratio predicted by this model, i.e.
the line flux of the Cr\,K${\alpha}$ should represent 33$\%$ of the
Fe\,K${\alpha}$. This is not compatible with EW=265$^{+147}_{-145}$\,eV, 
and EW$<$230\,eV found respectively at 5.4\,keV and 6.4\,keV. 
Therefore iron spallation seems to be an unlikely explanation in this case,
 as also found for other similar line observations
 (e.g., NGC\,3516 and Mkn\,766).\\

If instead the line originates from an annulus on the disk, then 
the small inclination angle derived above does not appear to be consistent 
with the classification of this object as a type 1.8 Seyfert, as one 
might expect a larger inclination angle of about 45--60$^{\circ}$. 
However, Nayakshin \& Kazanas \cite{NK01}, 
Turner et al. (\cite{Tu02}) for NGC\,3516, and Dov\v{c}iak et al. (\cite{D04}) 
 have proposed that these features could be due
 to localized hotspots on the accretion disk surface 
 following its illumination by flares. In this case, the range of transverse 
velocity shifts would be much smaller than for an entire disk orbit, 
resulting in a relatively narrow emission line feature, even at lower 
inclination angles. 
Narrow red-shifted features could be produced with an inclination angle
 of up to about 30 degrees in this model 
(see Fig.~1 in Nayakshin \& Kasanas \cite{NK01}, and also 
 Fig.~3 in Dov\v{c}iak et al. \cite{D04}),
 while at higher inclination angles the line becomes blue-shifted above 6.4 keV,
 which is ruled out from the data.
The narrow width of the observed emission line 
would imply that the emitting region must be small and that it is 
observed for only a fraction of the entire orbit. 
Indeed the orbital timescale of matter co-rotating 
along a circular trajectory assuming r=10$R_{\rm g}$ is between 
 2.7--27 hr for a black mass of 10$^{7}$\,M$_{\odot}$
 and 10$^{8}$\,M$_{\odot}$, respectively, 
 probably larger than the exposure time of the present observation. 
A longer exposure is therefore required to study possible
variations in the line profile as a function of phase of the spot. 

\indent The redshifted line emission could be also due to the signature of 
an inflow or outflow of matter from material close to the nucleus.  
Moving material that is not part of the disk structure would be sensitive
 to the effects of strong gravity and may achieve velocities that are a
 significant fraction of the speed of light, causing significant
 displacement of emission lines from their rest energy.
 In the XMM-Newton spectrum of Mrk\,766, 
Turner et al. (\cite{Tu04}) found an energy shift from 5.60 to 
5.75~keV in a line occurring over a few
 tens of ks. They proposed that such rapid energy 
shift could be interpreted
 as deceleration of an outflowing ejected blob of gas traveling to the escape
 velocity, where the material is viewed on the reverse side of the black hole 
(i.e. moving away from the observer). 
The energy of the line in ESO\,113-G010 
could indicate a (redshifted) fast moving 
medium at 0.16--0.23\,c, assuming a neutral iron
(6.4\,keV in its rest-frame) or a H-like iron (7\,keV in its rest-frame) 
respectively.  \\

 Up to now mainly red-shifted narrow lines 
 (except for the NLS1 RX J0136.9-3510, Ghosh et al. \cite{G04})
 have been observed in Seyfert 1 galaxies. 
For the hotspot model, this could be explained by the fact that 
red-shifted lines are predicted for very small 
 disk inclination (up to about 30 degree), 
 which is compatible with the Unified Scheme (Antonucci \cite{An93}). 
For the blob model, this could be explained also by a geometry
effect. The outflow could be optically thick on the near side of the flow,
 so one only sees emission from the reverse side. 
 However given the statistical quality of this dataset at energies above 
the iron K line ($>7$\,keV), it is not possible to exclude a blue-shifted 
component to the line from an outflow, which has the same strength as the 
redshifted 5.4 keV line. For instance the 90\% confidence upper limit on 
the equivalent width of a line detected between 7--8 keV is $<340$\,eV, 
i.e. consistent with the line observed at 5.4 keV.  

The present observation of ESO\,113-G010 
is far too short (4\,ks) to study any shift of the line energy 
and measure any outflow or inflow of the material.  
Indeed within the current snapshot observation, three 
scenarios may be possible; 
(i) the line emitting matter may be accelerating due to 
gravity onto the accretion disk or black hole; 
(ii) the material may be ejected from the accretion disk and decelerate 
under gravity away from the observer or 
(iii) the line may originate from a rapidly variable 
hot-spot on the disk surface.  
 In case of a material accelerating or decelerating 
under gravity, we can use $\Delta v$/$\Delta t$ = GM/R$^{2}$ 
(where $\Delta t$ is the time under which the deceleration is observed 
and $R$ is the radius at which the material is situated at that time) to estimate 
the observed time of velocity change. 
For example, assuming a change of the energy line from 5.4\,keV to 5.6\,keV, 
and R$\sim$ 3.5 $\times$ 10$^{14}$\,cm (as for Mrk 766, Turner et al. \cite{Tu04}), 
we inferred $\Delta t\sim$ 90\,ks and $\Delta t\sim$ 9\,ks, assuming a black hole mass 
 of 10$^{7}$\,M$_{\odot}$ and 10$^{8}$\,M$_{\odot}$, respectively. 
 In the case of the hotspot scenario, both line energy and line intensity are predicted to
 vary with time on the orbital timescale (from 10\,ks up to 100\,ks if the black hole mass 
is 10$^{7}$\,M$_{\odot}$ and 10$^{8}$\,M$_{\odot}$, respectively), 
as shown in Fig.~1 in Nayakshin \& Kazanas (\cite{NK01})
 (see also Figs.~3 and 4 in Dov\v{c}iak et al. \cite{D04}). 
Therefore a much longer {\sl XMM-Newton} observation, of about 100\,ks in length,
 is required to be able to differentiate between these models. 

In this paper we have performed Monte Carlo simulations to 
calculate a realistic detection probability 
for the redshifted line at 5.4 keV. The Monte Carlo 
method takes into account that an iron line-like residual may be detected 
at any energy within the iron line band (defined as 4--7\,keV in this 
instance); in contrast the conventional F-test is likely to over-estimate 
the detection significance, as this does not account for the range 
of energies or bins where a redshifted line may be observed. 
Applying the Monte-Carlo method 
to ESO\,113-G010 results in a detection significance 
of 99\%, which is still just under the $3\sigma$ confidence level. 
Clearly, a longer observation will be needed 
to increase the detection significance of the feature to $>99$\% confidence. 
This is also the first time a Monte-Carlo probability estimate 
has been reported for one of the several previously published 
narrow redshifted line features in AGN, 
e.g., \object{NGC 3516} (Turner et al. \cite{Tu02}), 
\object{ESO 198-G24} (Guainazzi \cite{Gu03}, Bianchi et al. \cite{B04}), 
\object{NGC 7314} (Yaqoob et al. \cite{Y03}), 
\object{Mrk 766} (Turner et al. \cite{Tu04}), 
and  \object{IC\,4329A} (McKernan \& Yaqoob \cite{MY04}). It would 
therefore be desirable to re-evaluate earlier claims on redshifted line 
detections using more robust 
Monte-Carlo techniques, as well as analysing a wider 
sample of Seyfert galaxies to determine the frequency 
of the relativistically shifted lines in AGN.

\section*{Acknowledgments}
Based on observations obtained with the {\sl XMM-Newton}, an ESA science
mission with instruments and contributions directly funded by ESA
member states and the USA (NASA). 
The authors thank the anonymous referee for fruitful comments and suggestions. 
D.P. acknowledges grant support from an MPE fellowship.

\end{document}